\documentclass{article}
\begin{document}
\def\p {{\partial}}
\def\n {{\nu}}
\def\m {{\mu}}
\def\a {{\alpha}}
\def\bt {{\beta}}
\def\f {{\phi}}
\def\th {{\theta}}
\def\g {{\gamma}}
\def\eps {{\epsilon}}
\def\e {{\psi}}
\def\la {{\lambda}}
\def\na {{\nabla}}
\def\bn {\begin{eqnarray}}
\def\en {\end{eqnarray}}
\title{Canonical path integral quantization of Einstein's gravitational
field}
\maketitle
\begin{center}
\author{S.I. MUSLIH\\Dept. of Physics\\ Al-Azhar university\\
Gaza, Palestine}
\end{center}

\begin{abstract}
The connection between the canonical and the path integral
formulations of Einstein's gravitational field is discussed using
the Hamilton Jacobi method. Unlike conventional methods, its shown
that our path integral method leads to obtain the measure of
integration with no $\delta$- functions, no need to fix any gauge
and so no ambiguous determinants will appear.
\end{abstract}
\section{Introduction}

A Lagrangian $L(q_i,\dot{q_i},t)$, is called regular if the rank
of the Hessian matrix
\begin{equation}
A_{ij}= \frac{\p^2 L}{\p \dot{q_i}\p \dot{q_j}},
\end{equation}
is $n$. On the other hand the Lagrangian is called singular if
the rank of the Hessian is less than $n$ e.g. $(n-r)$, $r > 0$.
Systems with this property are equivalently called "singular
Lagrangians", "constrained systems" or "singular systems".

Studies on singular systems started around $1950's$. Dirac [1,2],
initiated the well known method to investigate the Hamiltonian
formulation of constrained systems. His formulation became the
fundamental tool for the study of classical systems of particles
and fields. Bergman [3] and his collaborates work the
relationship between invariance principle and constraints in
field theories. Their efforts are to quantize Einstein's theory
of gravitation since this theory is a singular due to its general
covariance.

The study of Einstein's theory of gravitation using Dirac's and
Faddeev's methods has been widely investigated by many authors
[4-13]. In this paper we would like to obtain the path integral
quantization of Einstein's theory of gravitation using the
canonical path integral method [14-18].
\section{A brief review on the canonical path integral method}

The canonical formulation [19-22] gives the set of Hamilton-Jacobi
partial differential equations (HJPDE) as \bn
&&H^{'}_{\a}(t_{\bt}, q_a, \frac{\p S}{\p q_a},\frac{\p S}{\p
t_a}) =0,\nonumber\\&&\a, \bt=0,n-r+1,...,n, a=1,...,n-r,\en where
\begin{equation}
H^{'}_{\a}=H_{\a}(t_{\bt}, q_a, p_a) + p_{\a},
\end{equation}
and $H_{0}$ is defined as
\bn
 &&H_{0}= p_{a}W_{a}+ p_{\m} \dot{q_{a}}|_{p_{\n}=-H_{\n}}-
L(t, q_i, \dot{q_{\n}},
\dot{q_{a}}=W_a),\nonumber\\&&\m,~\n=n-r+1,...,n. \en

The equations of motion are obtained as total differential
equations in many variables as follows:

\bn
 dq_a=&&\frac{\p H^{'}_{\a}}{\p p_a}dt_{\a};\\
 dp_a=&& -\frac{\p H^{'}_{\a}}{\p q_a}dt_{\a};\\
dp_{\bt}=&& -\frac{\p H^{'}_{\a}}{\p t_{\bt}}dt_{\a};\\
 dZ=&&(-H_{\a}+ p_a \frac{\p
H^{'}_{\a}}{\p p_a})dt_{\a};\\
&&\a, \bt=0,n-r+1,...,n, a=1,...,n-r\nonumber \en where
$Z=S(t_{\a};q_a)$. The set of equations (5-8) is integrable [19]
if \bn
dH^{'}_{0}=&&0,\\
dH^{'}_{\m}=&&0,  \m=n-p+1,...,n, \en or in equivalent form
\begin{equation}
[H^{'}_{\a},\;\;H^{'}_{\bt}]=0\;\;\forall\; \a,\bt.
\end{equation}
Equations of motion reveal the fact that the Hamiltonians
$H^{'}_{\a}$ are considered as the infinitesimal generators of
canonical transformations given by parameters $t_{\a}$ and the set
of canonical phase-space coordinates $q_a$ and $p_a$ is obtained
as functions of $t_{\a}$, besides the canonical action integral is
obtained in terms of the canonical coordinates. In this case, the
path integral representation may be written as [14-18]

\bn
D({q'}_a,{t'}_{\a};q_a,t_{\a})=&&\int_{q_a}^{{q'}_a}~Dq^{a}~Dp^{a}\times
\nonumber\\&&\exp i \{\int_{t_{\a}}^{{t'}_{\a}}[-H_{\a}+
p_a\frac{\p H^{'}_{\a}}{\p
p_a}]dt_{\a}\},\nonumber\\&&a=1,...,n-r, \a=0,n-r+1,...,n. \en

Now we will study the path integral quantization of Einstein's
gravitational theory considering the method given in section 2.
\section{An example}

Let us consider the Lagrangian density of Einstein's
gravitational field as [4,5]
\begin{equation}
{\cal L} =N^{\bot} g^{\frac{1}{2}}(R + K_{ij} K^{ij} -k^{2}),
\end{equation}
 where $ g_{\m\n}$ are the metric and $R$ is the Riemann curvature scalar. The four functions $N_{\m}$ are treated
 as position variables, which are hidden in the extrinsic curvature
\begin{equation}
K_{ij}=\frac{1}{2N^{\bot}}(N_{i}|_{j} +N_{j}|_{i}-g_{ij,0}).
\end{equation}
The canonical momenta conjugated to $N^{\m}$ are

\begin{equation}
p_{\m}=\frac{\p {\cal L}}{\p(\p_{0}N^{\m})}=0,
\end{equation}
those conjugated to $g_{ij}$ are
\begin{equation}
{\pi}^{ij}=\frac{\p {\cal L}}{\p(\p_{0}g_{ij})}=-
g^{\frac{1}{2}}( K^{ij} -k g^{ij}).
\end{equation}
Taking the trace on both sides of relation (16), one gets
\begin{equation}
\pi=\pi_{i}^{i}=2g^{\frac{1}{2}}k.
\end{equation}
Hence, equation (16) can be solved for the $k^{ij}$ as
\begin{equation}
k^{ij}= -g^{\frac{-1}{2}}(\pi^{ij}- \frac{1}{2} \pi g^{ij}),
\end{equation}
in this case the 'velocities" $ g_{ij,0}$ can be expressed in
terms of the momenta $\pi^{ij}$ as
\begin{equation}
g_{ij,0}=-2g^{\frac{-1}{2}}N^{\bot}(\pi^{ij}- \frac{1}{2} \pi
g^{ij})- N_{i}|_{j} - N_{j}|_{i}.
\end{equation}
The canonical Hamiltonian density takes the form
\begin{equation}
{\cal H}_{0}= p_{\m}{\p_{0}}N^{\m} + \pi^{ij}{\p_{0}}g_{ij} -
{\cal L}.
\end{equation}
Making use of the primary constraint (15) and the expressible
velocities $g_{ij,0}$, we have
\begin{equation}
{\cal H}_{0}= 2\pi^{ij}N_{i}|_{j}-
g^{\frac{-1}{2}}N^{\bot}(\frac{1}{2} \pi^{2}-\pi^{ij} \pi_{ij} +
R g).
\end{equation}
After partial integration and neglecting the surface term, the
total canonical Hamiltonian can be expressed as
\begin{equation}
H_{0}= \int d^{3}x (N^{\bot} {\cal H}_{\bot} + N^{i} {\cal
H}_{i}),
\end{equation}
where

\bn {\cal H}_{\bot}=&&g^{\frac{-1}{2}}(\pi^{ij}
\pi_{ij}-\frac{1}{2} \pi^{2})-R g^{\frac{1}{2}},\\
{\cal H}_{i})=&&-2{\pi_{i}^{j}}|_{i}.
\en

Starting from the Hamiltonian (22) and making use of (15), the
canonical method [19-22] lead us to obtain the set of Hamilton
Jacobi partial differential equations as

\bn H'_{0}=&& p_{0} + H_{0}=0;\;p_{0}=\frac{\p S}{\p \tau},\\
H'=&&p_{\m}=0;\;\;\;\;\;p_{\m}=\frac{\p S}{\p N^{\m}}. \en Thus
one calculate the total differential equations as

\bn dg_{ij}=&&\frac{\p H'_{0}}{\p \pi^{ij}}d\tau + \frac{\p H'}{\p
\pi^{ij}}dN^{\m}=\frac{\p H'_{0}}{\p \pi^{ij}}d\tau,\\
d\pi^{ij}=&&-\frac{\p H'_{0}}{\p g_{ij}}d\tau - \frac{\p H'}{\p
g_{ij}}dN^{\m}=-\frac{\p H'_{0}}{\p g_{ij}}d\tau,\\
dp_{\m}=&&-\frac{\p H'_{0}}{\p N^{\m}}d\tau - \frac{\p H'}{\p
N^{\m}}dN^{\m}=-H_{\m}d\tau,\\
dp_{0}=&&-\frac{\p H'_{0}}{\p {\tau}}d\tau - \frac{\p H'}{\p
{\tau}}d{\tau}=0. \en

To check whether this set is integrable or not, one should
consider the total variations of (25) and (26). In fact, the total
variation of $H'_{0}$ leads to the conditions

\bn H_{1}=&& {\cal H}_{\bot}=g^{\frac{-1}{2}}(\pi^{ij}
\pi_{ij}-\frac{1}{2} \pi^{2})-R g^{\frac{1}{2}},\\
H_{2}=&&{\cal H}_{i})=-2{\pi_{i}^{j}}|_{i}. \en Since $H_{1}$ and
$H_{2}$ are not identically zero, we consider them as new
constraints, and one should consider the total variations of
$H_{1}$ and $H_{2}$ too. Calculations show that they are no
further constraints arise.

The set of equations (27-30) is integrable. Hence, the canonical
phase space coordinates $g^{ij}$ and $\pi^{ij}$ are obtained in
terms of independent parameters $\tau$ and $N^{\m}$. In this case
the path integral representation for this system is calculate as

\bn D({g^{ij}}', {\tau}',{N^{\bot}}', {N^{i}}' ;g^{ij},
\tau,N^{\bot}, N^{i})=&&\int \prod ~Dg^{ij}~D\pi^{ij}\times
\nonumber\\&&\exp i
\{\int_{\tau}^{\tau'}d^{3}x[N^{\bot}(g^{\frac{-1}{2}}(\pi^{ij}
\pi_{ij}-\frac{1}{2} \pi^{2})-R g^{\frac{1}{2}})\nonumber\\&&
-2N^{i}{\pi_{i}^{j}}|_{i}+ \pi^{ij}g_{ij,0}]d\tau \}. \en The
path integral representation (33) is an integration over the
canonical phase-space coordinates $g^{ij}$ and $\pi^{ij}$.

\section{ Conclusion}

We have obtained the canonical path integral formulation of
Einstein's theory of gravitation. This treatment leads us to the
equation of motion as total differential equations in many
variables, which require the investigation of integrability
conditions.

The Einstein's  gravitation system is integrable, $H'_{0}$ and
$H'$ can be interpreted as infinitesimal generators of canonical
transformations given by parameters $\tau$ and
$(N^{\m}=(N^{\bot}, N^{i}))$ respectively. Although $N^{\m}$ are
introduced as coordinates in the Lagrangian, the presence of
constraints and the integrability conditions force us to treat
them as parameters like $\tau$. In this case the path integral is
obtained as an integration over the canonical phase-space
coordinates $g^{ij}$ and $\pi^{ij}$. Other treatments [6-11] need
gauge fixing conditions to obtain the path integral over the
canonical variables.

An important point to specified here, is that for the other
conventional methods, there is no well defined procedure to
obtain the path integral amplitude for Einstein's theory of
gravitation. A formal expression for the amplitude may be written
as

\begin{equation}
F=\int dM(g_{\m\n}) \exp i S(g_{\m\n}).
\end{equation}
For the measure $dM$, starting with different assumptions,
different authors got different results. For example, Faddeev and
Popov [8], following Faddeev's method [12,13] find
\begin{equation}
dM_{FD}= \prod_{x} (-g)^{\frac{5}{2}} \prod_{\m\leq\n}dg^{\m\n},
\end{equation}
while Fradkin and vilkovisky [9-11] claim that
\begin{equation}
dM_{FV}= \prod_{x} (-g)^{\frac{7}{2}}g^{00}
\prod_{\m\leq\n}dg^{\m\n}.
\end{equation}
Besides, in reference [6] it is shown that, the local measure
emerging from canonical quantization of Einstein's theory of
gravitation, may in principle be omitted if the regularization is
properly used.

However the problems which arise naturally from identifying the
measure do not occur if our canonical path integral method is
used. Besides it is obvious that one dose not need to fix any
gauge if the canonical path integral method is used. All is
needed the set of the Hamilton Jacobi partial differential
equations and the set of the equations of motion. Then one should
tests whether these equations are integrable or not. If the
integrability conditions are not satisfied identically, then the
total variation of them should be introduced as new constraints
of the theory. Repeating this procedure as many times as needed
one may obtain a set of conditions. The number of independent
parameters of the theory is determined directly, without imposing
any gauge fixing conditions by this set.

\end{document}